\documentclass[useAMS,usenatbib]{mn2e}
\usepackage{amssymb, amsmath, epsfig}
\newcommand{\bfx}{{\boldsymbol{x}}}

\newcommand{\GeV}{{\rm GeV}}

\newcommand{\Msun}{{\rm M}_{\odot}}

\begin{document}
\title{Testing the Dark Matter Annihilation Model for the WMAP Haze}
\author[M. McQuinn \& M. Zaldarriaga]{Matthew McQuinn$^1$\thanks{mmcquinn@berkeley.edu} and Matias Zaldarriaga$^{2}$\\
$^{1}$ Einstein Fellow, University of California, Berkeley; Berkeley, CA\\
$^2$  Institute for Advanced Study; Princeton, NJ\\
}

\pubyear{2009} \volume{000} \pagerange{1}

\maketitle\label{firstpage}

\begin{abstract}
Analyses have found a ``haze'' of anomalous microwave emission surrounding the Galactic Center in the WMAP sky maps.  A recent study using Fermi data detected a similar haze in the $\gamma$-ray.   Several studies have modeled these hazes as radiation from the leptonic byproducts of dark matter annihilations, and arguably no convincing astrophysical alternative has been suggested.  We discuss the characteristics of astrophysical cosmic ray sources that could potentially explain this microwave and $\gamma$-ray emission.  The most promising astrophysical scenarios involve cosmic ray sources that are clustered such that many fall within $\sim 1$~kpc of the Galactic Center.  For example, we show that several hundred Galactic Center supernovae in the last million years plus a diffusion-hardened electron spectrum may be consistent with present constraints on this emission.  Alternatively, it could be due to a burst of activity probably associated with Sagittarius A$^*$ occurring $\sim 1$~Myr ago and producing $>10^{51}$~erg in cosmic ray electrons.  Different models predict contrasting trends for the spectral index of the microwave and $\gamma$-ray spectrum as a function of angle from the Galactic Center that we show should be robust to cosmic ray propagation uncertainties.  In particular, if the haze is from dark matter annihilations, it should have a very hard microwave and $\gamma$-ray spectrum for which the spectral shape does not change significantly with angle, which we argue would be difficult to achieve with any astrophysical mechanism.  Observations with the Planck and Fermi satellites can distinguish between viable haze models using these signatures.
\end{abstract}

\begin{keywords}
cosmology theory, dark matter, microwaves, diffuse radiation
\end{keywords}

\section{Introduction}

Using data from the Wilkinson Anisotropy Probe (WMAP), \citet{finkbeiner04a} discovered an approximately circular excess of diffuse emission centered around the
Galactic Center of radial extent $\sim 20^\circ$ and luminosity $L_{\rm haze} \sim
10^{37} \; {\rm erg \, s^{-1}}$ between $20$ and $60$ GHz.
Nicknamed the ``WMAP haze,'' this radiation was the residual emission after \citet{finkbeiner04a} subtracted templates for Galactic H$\alpha$, synchrotron, and dust emission from the WMAP foreground maps.  \citet{finkbeiner04a} argued that the spectrum of the haze was too hard to be
due to synchrotron from cosmic ray electrons produced in supernovae.  Therefore, this emission was initially attributed to free-free
emission from $\gtrsim  10^5$ K gas \citep{finkbeiner04a}.  However, in a follow-up study \citet{finkbeiner04} argued that the lack of diffuse X-ray emission from the Galactic Center ruled out this hypothesis.  If the haze is not due to free-free emission, it must be synchrotron.  Interestingly, \citet{finkbeiner04} showed that the energetics and profile of the haze emission could be explained by synchrotron emission from the byproducts of dark matter annihilations for a generic weakly interacting $\sim 100$ GeV particle with a standard dark matter halo profile.   This idea has been pursued further in several subsequent studies \citep{hooper07, hoopergray, zhang08, cholis08, cumberbatch09,  harding09}. 

However, cosmic rays from astrophysical sources may also be able to create the haze emission. The total energy in synchrotron electrons required to produce the haze is
\begin{equation}
E_{\rm syn}^{\rm haze} \sim 10^{51}\, \left(\frac{L_{\rm haze}}{10^{37} {\rm erg \, s^{-1}}} \right) \, \left(\frac{10 \, {\rm \mu G}}{B} \right)^{\frac{3}{2}}  \left(\frac{50 \,  {\rm GHz}}{\nu}\right)^{\frac{1}{2}} ~{\rm erg}. 
\label{eqn:energy}
\end{equation}
Equation (\ref{eqn:energy}) is derived by equating $L_{\rm haze}$ with the synchrotron loss rate for an electron that emits at $\nu$ in a magnetic field of strength $B$, noting the relationship between electron energy $E$ and the synchrotron peak frequency $\nu$:
\begin{equation}
E \approx  30  \; \left( \frac{\nu}{50 \; {\rm GHz}} \right)^{1/2} \; \left(\frac{B_\perp}{10 \; {\rm \mu G}}\right)^{-{1}/{2}} ~ {\rm GeV},  
\end{equation}
where $B_\perp$ is the magnetic field component orthogonal to the particle motion.
For comparison, $\sim 10^{51}~$erg is roughly equal to the kinetic energy of a single supernova explosion.   This energy must be output within the time it takes an electron to radiate its energy via synchrotron and inverse Compton or
\begin{equation}
\Delta t = 1.0 \;  \left( \frac{B_\perp}{10 \; {\rm \mu G}} \; \frac{50 \, {\rm GHz}}{\nu} \right)^{{1}/{2}} \left( \frac{U_{\rm r} + U_{\rm B}}{10 \, {\rm eV \,  cm^{-3} }} \right)^{-1}  ~ {\rm Myr},
\label{eqn:cooling}
\end{equation}
where $U_{\rm r}$ and $U_{\rm B}$ are respectively the energy densities in radiation and magnetic fields.

While it is conceivable that astrophysical sources could satisfy these energetics, the difficulties with creating the haze with astrophysical sources are threefold: 
\begin{enumerate}
\item \citet{dobler08} find the haze spectrum in the microwave to be hard, measuring a power-law electron spectrum with index $\gamma \approx 2$.  Such a spectrum is much harder than the galactic synchrotron towards other directions (where $\gamma \approx 3$). 
\item The distribution of known galactic cosmic ray sources appears to not be sufficiently concentrated toward the Galactic Center to create the haze \citep{finkbeiner04, dobler08, zhang08}.  The distribution of supernovae and pulsars is estimated to peak at $\approx 4$ kpc from
the Galactic Center, corresponding to an angular separation of $\approx 30^\circ$ \citep{lorimer06}.  In contrast, the haze intensity is strongly increasing with decreasing angle at $<20^\circ$ from the Galactic Center \citep{dobler08}.
\item By construction, a similar excess towards the Galactic Center is not present at $408$ MHz.  If the haze is from synchrotron, this suggests that the source of the haze does not contribute significantly to the population of cosmic ray electrons at $\sim 5$~GeV.  In addition, the sky at $408~$MHz is dominated by emission from the galactic disk, without a significant enhancement in the haze region.
\end{enumerate}

No study that has proposed an astrophysical explanation for the haze has demonstrated that it can satisfy all of these criteria.  Pulsars are the primary astrophysical solution that has been discussed in the literature \citep{zhang08, kaplinghat09, harding09}.  We argue here that a disk population of pulsars likely cannot satisfy the third criterion.  

If dark matter annihilations from the Milky Way's hypothetical dark matter density cusp are the source of the haze, \citet{finkbeiner04} found that these three criteria can easily be satisfied.  Furthermore, for standard dark matter models, large clumping factors or boost factors in the annihilation cross section above theoretically preferred values may not be required to explain this radiation \citep{finkbeiner04, hooper07}.  If large boost factors are not required, this would be in contrast to other recent astrophysical anomalies that have been interpreted as arising from dark matter annihilations.

Recently, a $\gamma$-ray ``haze'' was detected by \citet{doblerFermi} using data from the Fermi Space Telescope, which has a similar spatial morphology to the WMAP haze.  This haze was interpreted as inverse Compton emission off of starlight from the same hard population of electrons that generate the haze in the microwave \citep{doblerFermi}.  The $\gamma$-ray haze may be another difficulty for astrophysical models of the haze that likely can be accommodated in dark matter models \citep{cholis09}.  However, we argue that present measurements of this excess in the $\gamma$-ray do not yet provide a stringent test for models of the haze.

The bulk of this paper investigates different explanations for the haze emission and their observational signatures.  Section \ref{sec:cosmic_rays} summarizes the standard diffusive model for galactic cosmic ray propagation.  Section \ref{sec:sources} details several candidate sources for the haze.  Section \ref{sec:results} contrasts the microwave properties of the candidate haze models, and Section \ref{sec:IC} contrasts these models' inverse Compton signatures.  Throughout this article we refer to the radiating particles as ``electrons'', even though some fraction may be positrons.


\section{Galactic Cosmic Ray Propagation}
\label{sec:cosmic_rays}

Cosmic ray electrons travel through the Milky Way halo, scattering off of small-scale magnetic inhomogeneities.  While there is significant uncertainty in the character of cosmic ray trajectories, the path of cosmic ray electrons is typically approximated as a random walk because the distance a cosmic ray electron travels in one direction (approximately the Larmor radius or $10^{13} \, [E/30\; \GeV]\,[B_\perp/10 \, \mu G]^{-1}$~cm) is much smaller than the size of the galaxy and its radio halo.  In this approximation, the electron number density $n$ evolves according to 
\begin{equation}
\frac{\partial n}{\partial t} = \overrightarrow{\nabla} \cdot \left( K(E, \bfx) \, \overrightarrow{\nabla} n \right) +  \frac{\partial}{\partial E} \left(B(E, \bfx) \, n \right) + Q(E, \bfx), 
\label{eqn:diffeqn}
\end{equation} 
where $K$ and $B$ are diffusion coefficients, $Q$ is the source term, and $\bfx$ represents the galactic location (e.g. \citealt{berezinskii}).\footnote{Equation (\ref{eqn:diffeqn}) assumes that the diffusion is isotropic and that re-acceleration within the interstellar medium is unimportant.  Re-acceleration is thought to be unimportant at $E$ relevant to this study (e.g., \citealt{strong07}).  Cosmic ray data constrains re-acceleration of nuclei with an energy per nuclear charge of $>1$~GeV to be negligible, and re-acceleration should be even less relevant for electrons at these energies owing to their shorter lifetime \citep{berezinskii}.}

Under the simplistic assumption that the diffusion parameters are spatially independent, the number of cosmic ray electrons at distance $r$ away from a burst that occurred a time $\Delta t$ ago is (e.g., \citealt{berezinskii})
\begin{equation}
G(r , |\,\lambda) = \left[2 \pi \lambda^2 \right]^{-3/2} \, \exp \left( - \frac{r^2}{2 \lambda^2} \right), 
\label{eqn:green_func}
\end{equation}
where the diffusion length $\lambda$ is given by 
\begin{eqnarray}
\lambda^2 &=& 2 \, \int_E^{E_0} dE \, K(E)\, B(E)^{-1}, \nonumber \\
&=& \frac{2 \, K_0 \, \left[ 1 - \left(1 - A E \Delta t \right)^{1- \delta} \right]}{A \left(1 - \delta \right) E^{1- \delta}},
\label{eqn:lambda}
\end{eqnarray}
and $E_0 = E/(1 - A \, E \, \Delta t)$ is the input energy.  We have adopted the standard parameterizations $B(E) = A \, E^2$ (the scaling for inverse Compton and synchrotron losses) and $K(E) = K_0 (E/ 1 \, {\rm GeV})^\delta$, where $\delta > 0$ (such that more energetic electrons diffuse faster).  Also relevant is the electron distribution for a burst with an input spectrum of $Q_0 E^{-\gamma_{\rm int}}$ rather than a $\delta$-function at $E_0$.  In this case, the distribution of electrons at $E < (A \, \Delta t)^{-1}$ is 
\begin{equation}
N(r, E \, | \, \Delta t)  = Q_0 \, G(r \,|\, \lambda) \, (1-AE \Delta t)^{\gamma_{\rm int}-2} \, E^{-\gamma_{\rm int}}.
\label{eqn:distpl}
\end{equation}

In the limit that  $E \sim  E_0$ (equivalent to $A \, E \, \Delta t \ll 1$), equation (\ref{eqn:lambda}) reduces to 
\begin{equation}
\lambda = 2.3 \, \left(\frac{K_0}{10^{29} {\rm \; cm^2 \, s^{-1}}}\right)^{\frac{1}{2}}  \left(\frac{\Delta t}{10^6 \, {\rm yr}}\right)^{\frac{1}{2}}  \left(\frac{E}{30 {\rm \, GeV}} \right)^{0.3}  \, {\rm kpc}.
\label{eqn:lambda1}
\end{equation}
This limit is most relevant for characterizing the diffusion of an event in which all of the energy in cosmic rays was input at a single time.
In the opposite limit $A \, E \, \Delta t \rightarrow 1$, equation (\ref{eqn:lambda}) reduces to
\begin{equation}
\lambda =  3.6 \, \left(\frac{K_0/A}{10^{44} {\rm \, cm^2 \, GeV}}\right)^{\frac{1}{2}} \; \left( \frac{E}{30 \,{\rm GeV}} \right)^{-0.2} \; {\rm kpc}.
\label{eqn:lambda2}
\end{equation}
This second limit is the relevant limit for a source with a continuous output at energy $E_0 \gg E$ (which approximates the input if dark matter annihilations are the source; Section \ref{sec:source_DM}).   We find the the diffusion length has essentially the same scalings as in equation (\ref{eqn:lambda2}) for a continuous input of cosmic rays with a power-law spectrum.   Although, for an input power-law of $1.5 \lesssim \gamma_{\rm int} \lesssim 3$, the coefficient in front of equation (\ref{eqn:lambda2}) becomes $\approx 1.5-2.5$ rather than $3.6$.

Both equations (\ref{eqn:lambda1}) and (\ref{eqn:lambda2})
 are evaluated at $\delta = 0.6$, and the evaluated values for $\delta$, $K_0$, and $A$ are consistent with measurements as described below.  The diffusion distance in both of these equations is comparable to the radial extent of the haze or $\sim 3$~kpc.  Thus, it is possible that source(s) in the center of the Galaxy could be responsible for the haze.

A key difference between the continuous source case and the case in which a single event was the source is that presently less energetic electrons (which were released at an earlier time) have diffused further for a continuous source ($\lambda \sim E^{(\delta-1)/2}$ where $0.3 \lesssim \delta \lesssim 0.6$).  The opposite trend with energy holds for an event ($\lambda \sim E^{\delta/2}$). 
These different scalings arise because for an event $\lambda$ is set by how far electrons traveled in $\Delta t$, whereas for a continuous source $\lambda$ is set by the distance electrons of energy $E$ traveled prior to losing their last $\Delta E \sim E$ of energy.  In addition, the average spectral index of the electrons for an event is equal to $\gamma_{\rm int}$ for $E \ll (A \Delta t)^{-1}$, and there are zero electrons with $E > (A \Delta t)^{-1}$ (eqn. \ref{eqn:distpl}).  In contrast, for a continuous source the average index is equal to $\gamma_{\rm int} + 1$ for $\gamma_{\rm int} > 1$ and equal to $2$ otherwise.

To match cosmic ray observations, galactic cosmic ray models assume that cosmic rays escape freely at some distance $L_{\rm halo}$ above the disk \citep{strong07}.  One can generalize the solution to the diffusion equation, $G(r | \lambda)$, to satisfy such a boundary condition by the method of images with a convergent series of images.  Furthermore, the cosmic ray distribution for a spatially or temporally extended source can be derived via convolution with $G(r | \lambda)$.  We adopt this approach in most of our calculations.   The drawback to this approach is that the coefficients of the diffusion must be spatially constant.  We note that even the most sophisticated models for cosmic ray diffusion, such as the GALPROP code \citep{strong07}, assume that $K$ is independent of $\bfx$.

Unless stated otherwise, we use $B_\perp  = \sqrt{2/3} \,10~\mu$G and $A = 10^{-15} \, {\rm GeV^{-1} \, s^{-1}}$.  The latter is equivalent to $U_{\rm r} + U_{\rm B} \approx 10 \, {\rm eV \, cm^{-3}}$ and is several times the value at the solar circle.  This value for $B_\perp$ is consistent with the estimated value at $\approx 3$~kpc in Galactic radius \citep{beck01}.  
The parameter $A$ is degenerate with other diffusion parameters (both with $K$ and with $\Delta t$).  In addition, we adjust $K_0$ between $0.1$ and $2 ~\times10^{29} \, {\rm cm^2 \, s^{-1}} $ and $\delta$ between $0.3$ and $0.6$ in order to qualitatively reproduce the haze observations in the context of different source models.  The former value of $0.3$ is what is expected for Kolmogorov spectrum of magneto-hydrodynamical turbulence.  The latter value of $0.6$ is closer to what is expected for a Kraichnan-type
turbulent spectrum.   In addition, we set $L_{\rm halo} = 4$~kpc unless stated otherwise.  The standard diffusive cosmic ray models are able to fit present cosmic ray data with $\delta = 0.3-0.6$, $L_{\rm halo} = 3-5$ kpc, and $K_0 = 3-5\times 10^{28}~ {\rm cm^{2} \; s^{-1}}$ \citep{strong07}.  Although, larger $K_0$ can be accommodated if $L_{\rm halo}$ is also increased, and $K_0$ could be as much as a factor of $10$ larger in the radio halo than in the disk \citep{berezinskii}.\footnote{The constraints on the parameters of this standard diffusive model primarily derive from two measurements.  First, measurements of the ratio of different primary and secondary cosmic rays yield the column of protons that cosmic rays traverse (which relates to the distance traveled).  Second, the abundance of radioactive cosmic ray nuclei provide the lifetime of cosmic rays in the galactic diffusion zone.  However, it is likely that the cosmic ray diffusion parameters are different above the disk in the central $\lesssim 4$~kpc of the Galaxy from their effective value obtained from local measurements, but even the sign of how $K_0$ (for example) will differ from its locally-measured value is uncertain.}


\section{Potential Haze Sources}
\label{sec:sources}

To detect the haze, first \citet{dobler08} removed the Cosmic Microwave Background (CMB) temperature anisotropies from the WMAP sky maps, which can be much larger in amplitude than the foregrounds.  They took linear combinations of the different frequency channels that roughly add up to zero in brightness temperature to perform this removal.  There is uncertainty in how to do this correctly since the optimal linear combination depends on the properties of the foregrounds that one is measuring, and this uncertainty dominates the measurement errors on the haze profile.  To quantify this uncertainty, \citet{dobler08} used $6$ physically motivated estimators for the CMB (i.e., different linear combinations) to subtract the CMB anisotropies.  The shaded regions/points for the haze measurement that are included in some of the ensuing plots quantify the range of measured values from these different estimators.

Once the CMB anisotropies were removed, this yielded a map of the foregrounds.  Next, \citet{dobler08} simultaneously fit templates for galactic dust emission, free-free (using H$\alpha$ emission as a tracer), and ``soft'' synchrotron (using $408~$MHz data).  The haze was the residual emission that these templates did not account for.  Thermal dust is not a significant contaminant of the lowest WMAP frequency channels at which the haze detection is most secure (although spinning dust can contribute; \citealt{dobler08}), and H$\alpha$ is a robust tracer of free-free emission from $\sim 10^4$~K gas (provided that absorption by intervening dust is minimal).  In addition, direct and indirect emission from unresolved sources in the galactic bulge is also unlikely \citep{bandyopadhyay08}.   Thus, the most compelling models for the haze are either that it is free-free from gas with $T \gg 10^4$~K or that it is additional synchrotron radiation that is not captured in the soft-synchrotron template \citep{finkbeiner04a}.  This section discusses candidate sources for the haze that fall into these two categories.

\subsection{Free-Free Emission}

Soft X-ray maps of the inner galaxy show a similar spatial morphology to that of the haze \citep{snowden97}, which is suggestive that the haze owes to free-free emission from hot gas.  However, previous studies have shown that it is difficult to create the haze with free-free emission \citep{finkbeiner04, dobler08}.  In fact, if the haze also appears in the $\gamma$-ray, as suggested by \citet{doblerFermi}, it cannot be free-free.  We review the arguments for why this microwave emission most likely cannot arise from free-free processes.  

Firstly, \citet{dobler08} claimed that the radio spectral index of the haze (defined here as the power-law slope of the specific intensity) is inconsistent with free-free emission.  The radio spectral index of free-free radiation is $\approx 0.15$, whereas \citet{dobler08} measured a median radio spectral index for the haze of $\approx 0.4$ between the $23$ or $33$ and $41~$GHz channels.  However, there were still estimators in the \citet{dobler08} study that yielded a spectrum that was hard enough to be consistent with free-free emission.  Even if their claim is correct that the emission is too soft to be free-free, a component of the haze could arise from free-free emission, and this component plus a softer component could yield the observed haze spectrum.

If a significant component of the haze were due to free-free emission from $\lesssim 10^5~$K gas, the same gas would be visible in H$\alpha$ radiation \citep{finkbeiner04a}, and, if it were due to $\gtrsim 10^6~$K gas, it would be visible in the ROSAT soft X-ray maps of \citet{snowden97} \citep{finkbeiner04}.  Thus, only free-free emission from $\sim 10^{5}$--$10^6~$K gas is viable.  However, gas at such temperatures is typically thermally unstable \citep{field65}, and to produce the haze emission requires a large column density of ions at these temperatures ($1$ kpc with a density of $0.1 \, {\rm cm}^{-3}$). 

Energetically it is difficult for there to be a sufficient amount of gas in this range of temperatures to produce the haze. Since the free-free specific intensity is approximately independent of frequency for $h \nu < k_b T_{\rm g}$ (where $T_{\rm g}$ is the gas temperature), the free-free energy loss rate of such gas would be
\begin{equation}
\dot{E}_{\rm brem} \sim [\nu L_\nu]_{\rm haze} \frac{k_b T_{\rm g}}{h \, \nu_{\rm haze}} \sim  10^{56} \; \left(\frac{T_{\rm g}}{10^6 \; {\rm K}} \right) ~{\rm erg \, {\rm Myr}^{-1}}, 
\end{equation}
which requires the kinetic power of $\sim 10^5$ supernovae explosions per Myr for $T_{\rm g} = 10^6$~K gas (and larger powers are required if resonance line emission is also included).  It is unlikely that Galactic Center sources can supply these energetics. 

\begin{figure}
\begin{center}
\epsfig{file=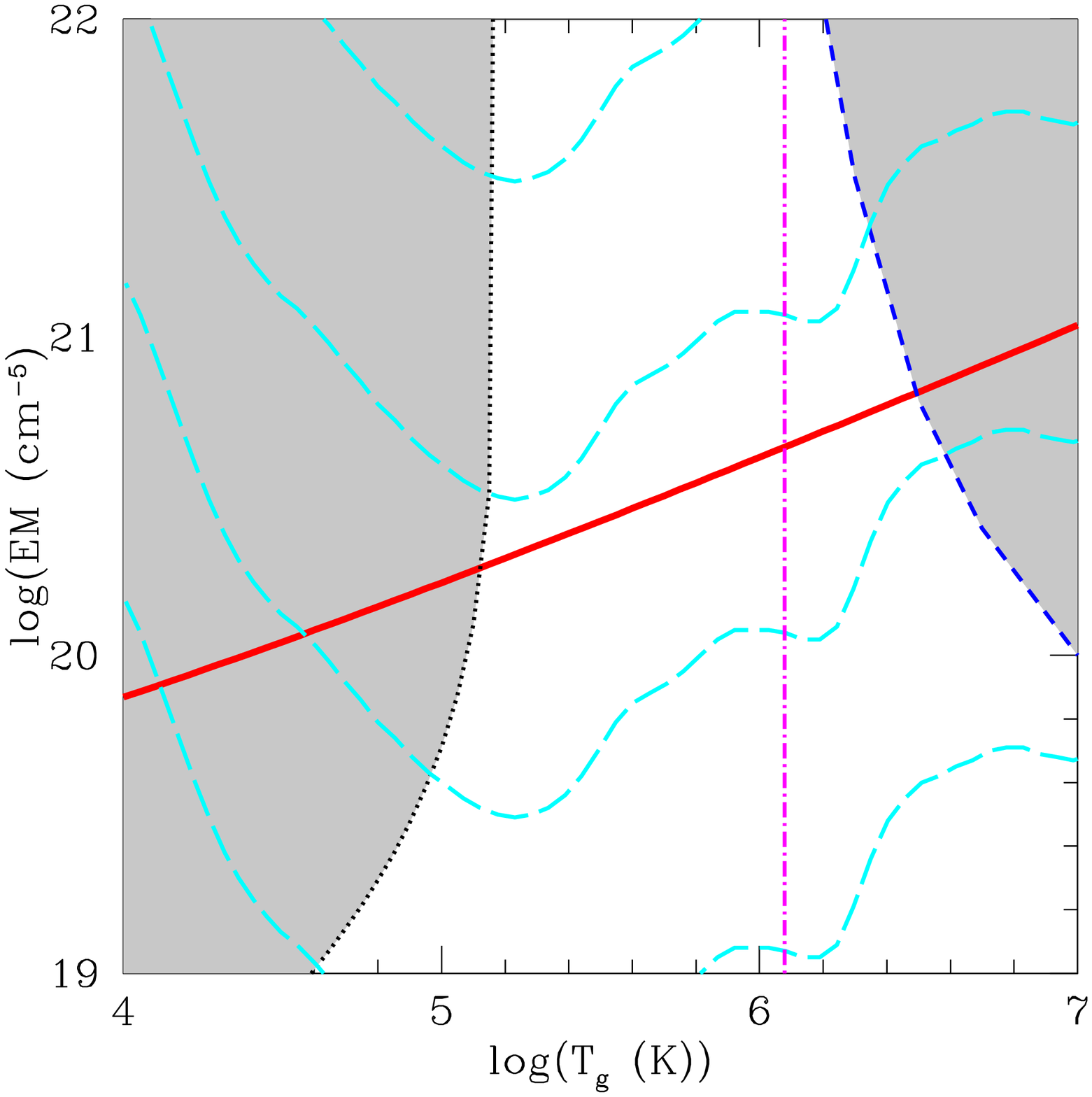, width=7.5cm}
\end{center}
\caption{Red solid curve represents the required $T_{\rm g}$ and EM if the haze owes to free-free emission (assuming $I_\nu = 2$~kJy~sr$^{-1}$, the haze value at ${b} = 10^\circ$).  The gray region on the left is excluded by H$\alpha$ measurements (with the limit $> 0.3$ Raleighs; \citealt{finkbeiner04a}) and on the right by X-ray measurements ($> 7$~Jy~sr$^{-1}$ at $750$~eV; \citealt{snowden97}).  The dashed curves are contours of constant energy input $\dot{E}$, using the approximate formula $\dot{E} =  2 \pi/3  \, \Lambda \, {\rm EM} \, \lambda_{\rm haze}^2$, where $\Lambda$ is the cooling rate density at solar metallicity \citep{sutherland93} and we set $\lambda_{\rm haze} = 2$~kpc --- the approximate extent of the haze.  The dashed curve that falls just above the solid red curve is for $10^{56}$ erg Myr$^{-1}$, and the other dashed curves represent $\dot{E}$ that are multiples of $10$ of this value.    The dot-dashed vertical line represents the temperature that balances the radial gravity of the disk for a circular velocity of $v_c = 100 \,$km s$^{-1}$ (roughly the Milky Way circular velocity at $2$~kpc) derived using $k \, T_g = 1.2 \, m_p v_c^2/2$.
\label{fig:freefree}}
\end{figure}

Figure \ref{fig:freefree} illustrates the constraints in the gas temperature ($T_g$) versus emission measure (EM) plane if free-free emission were the source of the haze, where EM is defined as the line-of-sight integral of the square of the electron number density.  The haze must fall on the solid line in this plane to yield the measured amplitude at ${b} = 10^\circ$.  The shaded region at the left is excluded by H$\alpha$ measurements and at the upper-right by soft X-ray measurements.\footnote{This region only accounts for the contribution of free-free emission to the X-ray flux.  The inclusion of the not insignificant emission from metal lines would shift this contour down and to the left, reducing further the allowed region.}  The long dashed curves are contours of constant energy loss for gas at the solar metallicity, with the curve just above the solid line representing $10^{56}$~erg Myr$^{-1}$, and the other curves step in factors of $10$ from this value.  In regions rightward of the dot-dashed vertical line, the gas cannot be gravitationally confined in the Galactic Plane, and, therefore, this parameter space is also excluded.   See the figure caption for more details.  Because of the required energetics, gas temperatures, and pressures, it is unlikely that the haze emission is due to free-free.

\subsection{Synchrotron Radiation from Cosmic Rays Produced by the Disk Population of Supernovae}
\label{ss:subtraction}

\citet{dobler08} used the Haslam $408$~MHz sky map as a template to subtract the ``soft'' synchrotron component of the foreground emission.  
This procedure assumed that the distribution of $\sim 3$~GeV electrons that is responsible for the $408$~MHz emission is the same as the distribution of $\sim 20$~GeV that is responsible for the soft synchrotron emission in the WMAP bands.  Even if the sources of the electrons at both energies are the same, this assumption can be violated by diffusion effects:  The diffusion length for a $20$~GeV electron is $1.4-1.9$ times smaller than a $3$~GeV electron for $0.3 < \delta < 0.6$.

The \citet{dobler08} analysis found that the soft synchrotron emission at $23$~GHz is roughly $\{1, 2, 3 \}$~kJy~sr$^{-1}$ at $\{25, 10, 5\}$ degrees in latitude and at longitude ${\it l} = 0$, whereas the haze is measured to be $\{1, 2, 5 \}$~kJy~sr$^{-1}$.  At fixed angle from the Galactic Center, but for ${\it l} \ne 0$, the soft synchrotron fraction is larger.  To create the correct amplitude for the haze at ${\it l} = 0$, a hardening of $\approx 0.25$ in the synchrotron spectral index ($\approx 0.5$ in $\gamma$) is required when extrapolating from $408$~MHz.

It would be difficult for diffusion effects from the disk population of supernovae to result in a change in $\gamma$ of $0.5$ in the inner $20^\circ$.   The distribution of galactic supernovae is measured to be a much smoother function of galactic longitude than the haze emission, with the favored model preferring a distribution that peaks at $4$~kpc or ${\it l} = 30^\circ$, and with few supernovae within $2$ kpc from the Galactic Center \citep{lorimer06}.   The supernova distribution is derived from emission measure determinations from hundreds of galactic pulsars and a model for the Milky Way interstellar electron density.  

However, the inferred distribution of pulsars depends strongly on modeling of the interstellar electron distribution.  \citet{lorimer06} showed that a more centrally concentrated distribution of pulsars can be obtained by assuming a less physically motivated, strongly peaked electron distribution towards the Galactic Center.  Yet, even in the most extreme case considered in \citet{lorimer06}, the inferred radial distribution is relatively flat in the central $\lesssim 5$~kpc, with $\approx 3\%$ of disk pulsars in the inner $1$~kpc.

\begin{figure}
\rotatebox{90}{\epsfig{file=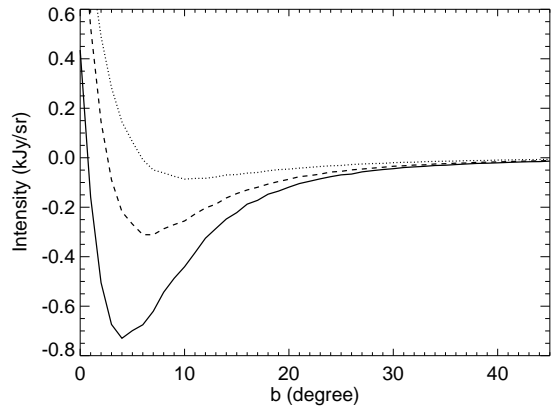, height=8cm}}
\caption{Residuals after subtraction of the fiducial GALPROP model prediction at $408$~MHz from its prediction at $23$~GHz, where the $408$~MHz prediction has been renormalized to minimize the square of the map differences.  The solid curve is for ${\it l} = 0$, the dashed for ${\it l} = 25^\circ$, and the dotted for ${\it l} = 40^\circ$. This procedure results in a small negative residual towards the Galactic Center at latitudes $b\sim 10^\circ$ for the standard distribution of cosmic ray sources.  Its amplitude is significantly smaller than the amplitude of the haze. 
\label{fig:subtaction}}
\end{figure}

For a source distribution that peaks $4$~kpc into the disk, or even a flat radial distribution in the inner several kpc, most of the galactic cosmic ray electrons must diffuse from their production sights a few kpc to reach the Galactic Center region.  Since less energetic electrons will have diffused further on average, diffusion causes the synchrotron emission from this population to soften toward the Galactic Center.  We used the GALPROP code to test this argument, as well as to test how well the $408$~MHz map can be used as template for the microwave synchrotron from supernova-accelerated electrons.  In particular, we used the default GALPROP parameters given in \citet{strong04}, as well as slight variants of these, to generate mock synchrotron sky maps.  GALPROP is a publicly available code that provides a numerical solution to equation~(\ref{eqn:diffeqn}) in a cylindrical region.\footnote{http://galprop.stanford.edu/web\_galprop/galprop\_home.html}  The default parameters are tuned to fit many different cosmic ray observations, using for the sources the standardly assumed distribution of supernovae.  In particular, the default model takes a radial distribution of supernovae in the disk that scales as $R^{0.4} \, \exp(-R/ 8 {\rm \, kpc})$.

We subtracted the $408$ MHz map generated with GALPROP from the $23$~GHz GALPROP map, minimizing the square of their difference in the region defined by $-45< {\it l} < 45^\circ$ and $0 < b < 45^\circ$ to emulate the method in \citet{dobler08}.  The results are shown in Figure \ref{fig:subtaction}.   This procedure resulted in a small negative signal toward the Galactic Center region at latitude $b \sim 10^\circ$ and longitude ${\it l} = 0$ (solid curve, Fig. \ref{fig:subtaction}).  This signal becomes less negative with increasing longitude (the dashed and dotted curves show ${\it l} = 25^\circ$ and ${\it l} = 40^\circ$).  This residual is an order of magnitude smaller in amplitude than the haze.\footnote{The divergent component at small angles in Figure \ref{fig:subtaction} owes to the fact that, for a continuous source population, the number density of electrons diverges as the distance to the source (which here is the disk of supernovae) decreases.  This divergence in Figure \ref{fig:subtaction} affects smaller angles than are relevant for the haze emission and should be ignored for this discussion.}   In addition, we find an almost identical result if we instead assume a flat radial distribution in the inner several kpc of the disk, with source density $\propto \exp(-R/ 8 {\rm ~kpc})$, analogous to the extreme case considered in \citet{lorimer06}.  While the GALPROP cosmic ray propagation model is simplistic, this small negative signal stems from the distribution of sources that is assumed and should be robust to the propagation model.

In general, a disk-like population of sources does not produce a spectrum where the spectral index varies significantly across the sky as long as the magnetic field is a weak function of radius and the solar circle is far from the radial edge of the diffusion region.  (These criteria appear to be roughly satisfied in the bulk of the Galaxy owing to the approximately constant spectral index that is observed across the bulk of the sky between the $408$~MHz maps and those in the microwave.)  To understand this, imagine different cylindrically symmetric diffusion zones that characterize different regimes for the electrons/magnetic fields around the disk of sources.  All sightlines from within this disk (that do not intersect the radial boundary) will penetrate each of these diffusion zones in the same proportion, resulting in the same spectral shape for the electron (and synchrotron) spectrum as a function of angle.  

In contrast, for a point cosmic ray source in the Galactic Center, the diffusion zones are more spherical, and sightlines will intersect these zones in different proportions.  The rest of this section investigates different point source models for the cosmic ray injection.

\subsection{Synchrotron Radiation from Cosmic Rays Produced by Supernovae in the Galactic Center}
\label{sec:population}\label{sec:SN}

We argued that the supernova population inferred from radio pulsar surveys likely cannot be the source of the haze emission.  
However, there is a population of undiscovered pulsars in the Galactic Center (indicating additional supernova activity) that has been missed by radio pulsar surveys and that are not included in galactic cosmic ray models. \citet{cordes97} demonstrated that previous pulsar surveys would not have identified pulsars in the central couple hundred parsecs because pulse broadening from dense gas in the Galactic Center washes out the pulses at the surveyed frequencies.

In addition, the observed star formation rate in the Galactic Center can produce enough supernovae to satisfy the haze energetics.  For $U_B + U_r = 10$~eV~cm$^{-3}$, $1000$ supernovae in the Galactic Center must have occurred in the last $10^6$~yr in order
to source the required $\sim 10^{51}$ erg in electrons if $\sim 0.1\%$ of the kinetic energy of each supernova is converted into GeV electrons that reach the ISM.  This percentile of $10^{51}$~erg (or $10^{48}$~erg) is the amount of energy that has been estimated to go into $E > 1$~GeV electrons from SN~$1006$ and also from observations of the intergalactic cosmic ray electron spectrum combined with estimates for the Milky Way supernova rate (see \citealt{kobayashi04} and references therein).\footnote{\citet{thompson06} derived a largely model-independent estimate of $\approx 10^{49}~$erg per supernova from the far infrared-radio correlation in starburst galaxies.}  One thousand is $\sim 10\%$ of the number of supernovae that occurred in the Galaxy within this period, which is a reasonable fraction since $10\%$ of the present galactic star formation is estimated to reside in the inner couple hundred parsec region \citep{figer08}.

A complementary estimate for the number of required supernovae can be derived by noting that the haze is comparable in amplitude to the emission from soft synchrotron (which is attributed to supernovae) within the inner $20^\circ$, and soft synchrotron emission of comparable amplitude covers $f_{\rm sky}$ of the sky, where $f_{\rm sky} \sim 0.1$.  Galactic electrons at these energies are expected to lose roughly half of their energy by synchrotron emission (and roughly half by inverse Compton).   Therefore, the fraction of Galactic supernovae required to produce the haze is $\sim (20^\circ)^2 /[4 \pi \times (180/\pi)^2 \times  f_{\rm sky}] \approx 0.1$.\footnote{These estimates assume that $\sim 30~$GeV electrons can escape from the central couple hundred parsec region.  There are some indications of milli-Gauss fields in this region \citep{plante95}, and, if this were the case, haze electrons would likely not be able to escape from this region.  However, it is probable that milli-Gauss fields that have been detected are localized to bundles and that in most of the volume the magnetic fields are comparable to in the rest of the Galaxy \citep{uchida95}.}

\begin{figure}
\begin{center}
\epsfig{file=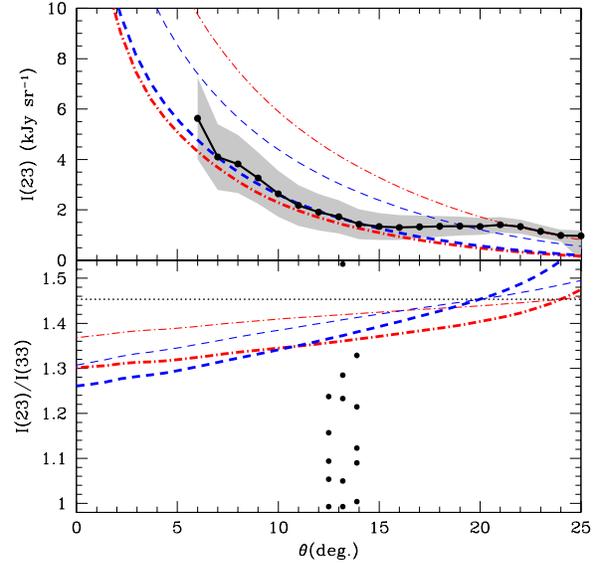, height=8cm}
\end{center}
\caption{Supernova models for the WMAP haze.  Top panel: The intensity profile as a function of angle from the Galactic Center for two supernova models described in the text (dot-dashed and dashed curves) as well as the measured intensity profile of the WMAP haze in the $23$~GHz channel (solid black curve with dots, with the grey region quantifying the current measurement uncertainty).  The thin curves are our models' raw intensity at $23$ GHz.  The thick curves are this minus their intensity at $408$~MHz scaled to $23$~GHz, assuming $I_\nu \sim \nu^{-1}$ to rescale the $408~$MHz map, emulating the template subtraction procedure in \citet{dobler08}.  Bottom panel:  The ratio of the intensity in the $23$ and $33$ GHz channels.  The curves show the same models as in the top panel.  The dotted horizontal curve is this ratio for $I_\nu \sim \nu^{-1}$ --- indicative of the galactic soft-synchrotron.  The filled circles represent the hardness values as quantified by the ratio of the total haze intensity in the $23$ and $33~$GHz channels for the $6$ different estimators to remove the primordial anisotropies in \citet{dobler08}.  The leftmost column of circles uses the power-law index inferred between the $23$ and $33$~GHz channels, the middle that between the $33$ and $41$, and the rightmost that between the $23$ and $41$.  \label{fig:sn}}
\end{figure}

If supernovae are the source of the haze, diffusion effects are required to harden the electron spectrum in the haze region.  Galactic Center supernovae can be treated as a continuous source population such that electrons that have diffused the furthest will have cooled the most (eqn. \ref{eqn:lambda2}).  Therefore, diffusion results in the synchrotron spectrum from Galactic Center supernovae softening as the distance from the Galactic Center increases.  At $r < \lambda$, the haze electrons should be hardened in this model by $0.6 < \Delta \gamma < 1.1$ for $0.3 < \delta < 0.6$ noting that $\lambda \sim E^{(\delta-1)/2}$ and $n(E', r)/n(E, r) \sim \lambda(E)^3/\lambda(E')^3$ as small $r$.  The thin curves in the left panel in Figure \ref{fig:sn} show the intensity profile for two simple supernova models; dashed: $K_0 = 2\times10^{29}$~cm$^2$ s$^{-1}$, $\gamma_{\rm int} = 2$, $\delta = 0.3$, and with input electron cosmic ray luminosity $\dot{E}_{\rm tot} =2\times10^{38}$~erg~s$^{-1}$ (equivalent to $6\times10^{51}$~erg~Myr$^{-1}$); dot-dashed: $K_0 = 1\times10^{29}$~cm$^2$ s$^{-1}$, $\gamma_{\rm int} = 2$, $\delta = 0.6$, and $\dot{E}_{\rm tot} =3\times10^{38}$~erg s$^{-1}$.  To calculate $\dot{E}_{\rm tot}$, the input spectrum of cosmic rays is assumed to extend between $1$~GeV and $1$~TeV.  Both models take a value for $K_0$ that is slightly higher than is typically used in cosmic ray diffusion calculations. (Although, $A$ might also be on the high side, and the quantity of importance for diffusion is $K_0/A$.)  The choice of $\gamma_{\rm int} = 2$ for supernovae is empirically motivated.\footnote{
Numerous observations have been made of the galactic population of electron cosmic rays at GeV energies, which are believed to be accelerated in supernova remnants.  The fiducial cosmic ray model used in \citet{strong04} and in most GALPROP calculations assumes that the injected supernova spectrum transitions over $\sim 2-20$~GeV from $\gamma_{\rm int} = 1.5$ at lower energies to $\gamma_{\rm int} = 2.5$ at higher energies, with an effective injection spectrum between $3$ and $20$~GeV of $\gamma_{\rm int} \approx 2$.  This extended transition is used to fit a break in the locally measured cosmic ray electron spectrum \citep{strong04}. (Although, at least some of this observed break may be explained by the increased importance of collisional losses at a couple GeV.)  Some cosmic ray models have assumed a single power-law of $\gamma_{\rm int} = 1.9$ across all relevant energies, and these models are able to fit much of the data \citep{strong00}.  Empirically, the spectral index of the soft synchrotron between the radio and microwave (and across different microwave bands) is measured to be $\gamma \approx 3$ on average, the anticipated spectrum after cooling if $\gamma_{\rm int} = 2$.}  

Both of these models are normalized to over-predict the measured haze intensity (the solid black curve with dots in the top panel in Fig. \ref{fig:sn}) for reasons detailed below.  Both models also predict a softer spectrum as quantified by the ratio of the WMAP intensity in the $23$ and $33$~GHz channels than the inferred estimates for this ratio from \citet{dobler08}.  The inferred estimates use the measured haze $\gamma$ between three permutations of the lowest three WMAP frequency channels plus six different estimators to remove the CMB anisotropies (the filled circles in the bottom panel of Fig. \ref{fig:sn}).\footnote{Unfortunately, present measurements of the haze are not able to constrain the hardness as a function of radius, which would be the ideal discriminator of the models presented in this paper.  What has been measured is the amplitude of the emission in the different frequency channels assuming a template with profile $I_\nu(\theta) \sim 1/\theta - 1/\theta_0$ for $5 < \theta < \theta_0 \equiv 45^\circ$.  The average intensity-weighted angle for this profile is $\langle \theta \rangle = 18^\circ$, which is the angle at which we plot the intensity ratios in the bottom panel of Figure \ref{fig:sn}.
}

The thick curves in Figure \ref{fig:sn} may be closer to what is actually measured than the thin curves.  These curves represent the intensity after subtracting the specific intensity $I_\nu$ of the corresponding model at $408$~MHz scaled to $23~$GHz assuming $I_\nu \propto \nu^{-1}$, the measured average radio spectral index for the Galactic soft synchrotron.
  This subtraction should be similar to how the haze is actually obtained.   The thick curves in the top panel in Figure \ref{fig:sn} are able to largely reproduce the profile of the haze emission at $23$~GHz for $\theta \lesssim 20^\circ$.  However, these haze models produce a spectral hardness that is only consistent with the couple estimators that yield the softest values for the haze hardness (bottom panel, Fig. \ref{fig:sn}).

There are two additional potential problems with the Galactic Center supernova explanation for the haze.  One problem is that to reproduce the amplitude of the haze at $23$~GHz requires an amplitude prior to subtracting the soft synchrotron that is $150-200\%$ of the amplitude of the haze in our two models in Figure \ref{fig:sn} (thick curves).  A value of $200\%$ is close to the total observed emission in the haze region at $23$~GHz, and such emission will significantly cut into the budget for the soft-synchrotron contribution from the disk population of supernovae.  The second problem is that the electrons in our two supernova models contribute to the emission at $408$~MHz at the $25-50\%$ level in the haze region.  However, there is no significant centralized emission in the inner $\sim 20^\circ$ in the $408$~MHz maps, at least at the $50\%$ level.   

One way to alleviate both of these potential issues is if the diffusion length were longer than we have assumed such that the haze spectrum from supernovae were even more hardened by diffusion  (Section \ref{sec:results}).  Another way is if there were a break in the injected electron spectrum of supernova between $5$~GHz and $20$~GHz such that the spectrum is harder at lower energies.  Such a break may even be required to fit local cosmic ray data.\footnotemark[\value{footnote}]   Such a break could increase the average diffusion length of electrons at $408$~MHz, making the emission in this band from the Galactic Center sources more diffuse.  However, this solution would also reduce the additional hardening our supernova models gain during the subtraction of the soft synchrotron.

\subsubsection{Pulsars Instead?}
\label{ss:pulsars}

Most previous astrophysical explanations for the haze have focused on pulsars \citep{zhang08, kaplinghat09, harding09}.  While it is not clear how efficiently pulsars can produce interstellar cosmic ray electrons,  several studies have attributed recently discovered cosmic ray ``anomalies'' in the ATIC \citep{chang08} and Pamela data sets \citep{adriani09} to these objects (e.g., \citealt{profumo08, hooper08}).   This led to several studies investigating whether pulsars could also explain the WMAP haze anomaly.  However, the generation of the haze with pulsars may be incompatible with the explanation that the ATIC and Pamela anomalies owe to pulsars \citep{zhang08}.

Young pulsars ($\lesssim 10^4$~yr) have been conceived as a promising model for the haze because they are conjectured to supply a harder source of electrons and positrons at these energies than supernovae.  \citet{kaplinghat09} and \citet{harding09} argued that a disk population of young pulsars could be responsible for the haze if these pulsars output $\sim 10\%$ of their spin-down energy into interstellar cosmic ray electrons and have $\gamma_{\rm int} \approx 1.5$.  In addition, \citet{kaplinghat09} showed that varying the longitudinal distribution of the uncertain galactic magnetic field could produce a similar intensity profile to the haze even from a disk population of pulsars. 

However, previous analyzes that invoked pulsars to explain the haze largely neglected that, when the soft synchrotron is subtracted from the WMAP maps using the $408$~MHz data as a template, this procedure should remove much of the pulsar contribution in the microwave because young pulsars have essentially the same spatial distribution as supernovae.\footnote{\citet{kaplinghat09} suggested one mechanism that could decouple the spatial distribution in these bands:  If the dominant contribution to the emission at $408~$MHz owes to cosmic ray secondaries rather than primaries.  Local measurements of the positron fraction suggest that $\sim 10\%$ of the e+e- that contribute at $408~$MHz owe to secondaries.
Therefore, while possible since local measurements may not be representative of the galaxy, this hypothesis is not supported by present data.  It is also unclear how different the distribution of secondary sources, which should largely trace the gas distribution, will be from that of the primaries, which traces star formation (which to some extent traces the gas).}    Therefore, pulsar models also require a diffusion hardened electron spectrum in the Galactic Center to produce a haze-like synchrotron signature.  The same arguments in Section \ref{ss:subtraction} for why a disk population is unlikely to produce the haze also apply in this case.

\subsubsection{Supernovae with a Galactic Wind}

A wind would naturally harden the spectrum towards the Galactic Center by transporting cosmic rays that have lost the most energy the furthest.\footnote{The haze electrons themselves cannot drive a strong wind.  The pressure from the haze electrons is minute compared to that from the much more abundant cosmic ray protons.}     In the limit of zero diffusion and zero adiabatic losses, the electron spectrum of a wind should have a spectral index of $\gamma = \gamma_{\rm int}$ for $E \lesssim (A \, r/v)^{-1}$ and have a cutoff due to cooling above this energy, where $r$ is the distance from the Galactic Center and $v$ is the wind velocity.   For $U_B + U_r = 10$~eV~cm$^{-3}$, a wind can harden the electron spectrum over scales comparable to that of the haze if it can transport particles $\gtrsim 2$~kpc within $1$~Myr (the timescales for the electrons that would produce the haze to lose their energy).  To meeting these conditions would require a $2000$~km s$^{-1}$ wind.  

There may be evidence for a galactic wind in the Milky Way \citep{2003ApJ...582..246B, everett08}.  \citet{everett08} interpreted a ROSAT soft-X-ray excess with similar extent to the haze as indicating a wind, and their best fit models yield $v = 200-300$ km s$^{-1}$.  However, for  $v = 300$~km s$^{-1}$ and $U_B + U_r = 10$~eV~cm$^{-3}$, $30$~GeV electrons could only reach a radius of $0.3$~kpc --- much less than the extent of the haze.  A haze-like extent would require a significantly smaller value for $U_B + U_r$ and/or a much faster $v$.  In addition, these estimates ignore adiabatic losses, which would be present for any expanding outflow.  We conclude that it is unlikely that a wind alone can produce the haze (unless there is in-situ electron acceleration in the wind).

\subsubsection{Other Source Populations in the Galactic Center}
\label{ss:othercont}

The discussion in subsection 3.3 also applies to other populations of sources in the central $\sim 1~$kpc of the Galaxy.  Any such source population would have the same difficulty as supernovae of being consistent with measurements of the haze hardness unless they had an injection spectrum with $\gamma_{\rm int} \lesssim 2$.  For example, tidal disruptions by stars that fall close enough to the supermassive black hole are estimated to occur every $10^{-4}$ to $10^{-5}$~yr$^{-1}$, with kinetic energies of $10^{51-52}~$erg per event \citep{strubbe09}.  If a small fraction ($\sim 1\%$) of this energy goes into accelerating electrons, the energetics would be sufficient to produce the haze.  Secondly, the accretion rate onto Sagittarius~A$^*$ is estimated to be a few$\, \times 10^{-6}~M_{\odot}$ from winds off of nearby stars \citep{quataert99, genzel04}, and upper limits on this rate based on observations coupled with accretion models are $1-2$ orders of magnitude larger \citep{quataert99}.  If even a small fraction of this rest mass energy contributed to accelerating electrons (possibly via an associated jet), this low level of accretion would also be sufficient to power the haze.


\subsection{A Single Energetic Event in the Galactic Center}
\label{sec:source_ps}

The next possibility that we consider is whether a single explosive event could be responsible for the haze.  Such a transient must output $\gtrsim 10^{51}\; {\rm erg}$ (eqn. 1; and we find $\sim 10^{52}\; {\rm erg}$ in  Section \ref{sec:results}).  For this model to be different from that of a continuous source population,
this event must be sufficiently rare in order for a cooling break --- a chromatic break in the electron spectrum to a power-law index of $\gamma_{\rm int} +1$ from $\gamma_{\rm int}$ --- to
not appear in the microwave from the accumulation of electrons from previous events.  The absence of a cooling break at relevant energies also enables such an event to have a hard enough spectrum even with $\gamma_{\rm int} > 2$, in contrast to continuous models.  

For electrons to be present that radiate significant energy at frequency $\nu_{\rm max}$ and for a cooling break to not have developed at $\nu > \nu_{\rm min}$, such a blast must have occurred a time
\begin{equation}
\tau_{\rm blast} < 1.0 \;  \left( \frac{B_\perp}{10 \; {\rm \mu G}} \; \frac{50 \, {\rm GHz}}{\nu_{\rm max}} \right)^{{1}/{2}} \left( \frac{U_{\rm r} + U_{\rm B}}{10 \, {\rm eV \,  cm^{-3} }} \right)^{-1}  ~ {\rm Myr}
\label{eqn:tau_blast}
\end{equation}
ago.  In addition, the last prior energetic event happened
\begin{equation}
t  > 1.5 \left(\frac{\tau_{\rm blast, max}}{1 ~ {\rm Myr}}\right) \; \left(\frac{\nu_{\rm max}/\nu_{\rm min}}{2.5} \right)^{{1}/{2}} ~~~ {\rm Myr} 
\end{equation}
ago, were $\tau_{\rm blast, max}$ is the value that saturates
inequality (\ref{eqn:tau_blast}).  Events that occur every $\sim 1$~Myr will often
accommodate both inequalities. 

Electrons from an event that occurred $1$~Myr ago can diffuse several kpc (eqn. \ref{eqn:lambda1}), on the order of the radial extent of the haze.  The most energetic electrons diffuse the furthest for an event since $\lambda \sim E^{\delta/2}$, such that the electron spectrum will harden with radius (eqn. \ref{eqn:lambda1}).  In addition, such an event will likely not be visible in the Haslam maps because there are far fewer electrons at a few GeV if the explosion has $\gamma_{\rm int} < 3$ than the cumulative number of electrons from supernovae (for which $\gamma \approx 3$).

Possibilities to create the haze from a single event include episodes that mark the deaths of massive stars (such as a rare supernova) and activity by the supermassive black hole (SMBH) in the Galactic Center.  However, such an event would have to produce $\sim 10^{52}$ erg in $\sim 30$~GeV cosmic ray electrons (Section \ref{sec:results}), which is likely too large to be from a single stellar event.  Whereas, transient activity associated with the SMBH in the Galactic Center can more easily accommodate these energetics.   In fact, there is evidence for recent activity in the Galactic Center.  In the $10^4~M_{\odot}$ Arches cluster, which sits $30$~pc from Sagittarius A$^*$, massive stars that have ages $< 2$ Myr are noticeably absent, indicating that this cluster formed $\approx 2~$Myr ago.  In addition, two other young clusters near Sagittarius~A$^*$ are estimated to have twice this age, one of these being the Central Cluster in the inner $1~$pc \citep{figer08}.  It is conceivable that the formation of one of these clusters was accompanied by a short episode of accretion onto Sagittarius~A$^*$.  If the $4\times10^6~\Msun$ SMBH shined at less than the Eddington luminosity during a period of gas accretion, this period must have lasted $> 0.6 f^{-1}$~yr.  The parameter $f$ is the ratio of the energy that goes into accelerating cosmic ray electrons, taken to be $10^{52}$~erg, to the energy in electromagnetic radiation.  

One compelling candidate for sourcing such an occurrence is the tidal disruption of a small molecular cloud by Sagittarius-A$^*$.  Molecular clouds are estimated to be tidally disrupted by Sagittarius~A$^*$ every $\sim 10^7~$yr in the Milky Way \citep{sanders98}.   In fact, an explanation for the origin of the S-stars around Sagittarius-A$^*$ is that they were formed several Myr ago from the debris of such an event \citep{sanders98}.  The timescale for the activity associated with the tidal disruption of a molecular cloud by Sagittarius-A$^*$ is the time for the cloud to cross the black hole.  For a cloud size of $0.1~$pc and a relative velocity of $10~$km~s$^{-1}$ with respect to Sagittarius~A$^*$, the cloud crosses the SMBH in $10^5~$Myr, which is much less than $1~$Myr as required.  Secular activity of this sort may be responsible for the nuclear activity in Seyfert galaxies \citep{hopkinsSeyfert}.  However, it is also possible that much of the accretion associated with the tidal disruption of a small molecular would occur at small enough accretion rates such that the accretion disk would be in a radiatively inefficient state.  In this case, such phenomena would be more difficult to detect in other galaxies.




\subsection{Dark Matter}
\label{sec:source_DM}

Cosmic rays generated as the byproducts of dark matter annihilations is the most exciting proposal for the source of the haze.  With well-motivated assumptions for the distribution of dark matter and its properties, the total luminosity in dark matter annihilations from the Milky Way halo can be estimated:
\begin{equation}
L_{\rm ann}(E) \approx 2 \times 10^{37}  \; {\cal B} \, \left(\frac{c}{12}\right)^{2.0} \, \frac{M_{\rm MW}}{10^{12} \,M_{\odot}} \,  \frac{100 \, {\rm GeV}} {M_{\rm DM}} ~{\rm erg \; s^{-1}}.
\label{eqn:Lann}
\end{equation}
This expression assumes an NFW dark matter halo profile \citep{navarro96} with total dark matter mass $M_{\rm MW}$ and concentration parameter $c$.\footnote{We assume a flat $\Lambda$CDM cosmology with $\Omega_M = 0.27$ and $h = 0.7$ to calculate the $z=0$ virial radius in spherical collapse, which is necessary to define $c$.}  The scaling for $c$ in equation~(\ref{eqn:Lann}) is approximate and most valid around $c = 12$, the value that matches rotation curve data for $M_{\rm MW} = 10^{12}~M_\odot$ \citep{besla07}.  $M_{\rm MW} = 10^{12}~M_\odot$ is on the low side of Milky Way mass estimates, with most estimates resulting in a value that is a factor of $2$ larger (e.g. \citealt{sakamoto03, li08}).   $M_{\rm DM}$ is the dark matter particle mass, and popular theories for weakly interacting massive particle (WIMP) dark matter predict $M_{\rm DM} \sim 0.1-1$~TeV.  The parameter ${\cal B}$ is a boost (or suppression) factor that may arise from additional clumpiness on top of the smooth NFW profile or if the value of the WIMP velocity-averaged annihilation cross section $\langle \sigma \, v \rangle$ is different in the Milky Way than at freeze out (where $\langle \sigma \, v \rangle \approx 3 \times 10^{-26}~{\rm cm}^3 \;{\rm s}^{-1}$ is required to match the relic abundance).  Recent numerical work argues that the contribution to ${\cal B}$ from clumping is near unity inward of the solar circle \citep{springel08, kamionkowski10}, and standard WIMP models typically predict that the Galactic value of $\langle \sigma v \rangle$ will be comparable to its value at freeze out.  However, numerical simulations are far from resolving the smallest dark matter structures, and dark matter models with ${\cal B} \gg 1$ have been invoked by several recent studies to explain several perceived astrophysical anomalies (e.g., \citealt{arkani-hamed08}).  It is interesting that for standard values of the halo and dark matter properties, the dark matter annihilation luminosity (eqn. \ref{eqn:Lann}) is comparable to the $\sim 10^{37}$~erg~s$^{-1}$ required to produce the haze.

The synchrotron haze emission from dark matter annihilations depends on the parameterization of the Galaxy's dark matter profile, magnetic field structure, and distribution of interstellar radiation.   Because of these uncertainties, adding complications in the dark matter sector of our model is not warranted, and our dark matter models simply assume direct annihilations into pairs (a $\delta$-function injection at $E = M_{\rm DM}/2$).  Such models results in the average spectrum of all electrons having $\gamma = 2$ for $E < M_{\rm DM}/2$.  More physically motivated WIMP models (in which e+e- are produced after a particle cascade) generally yield electron spectra that are slightly softer, and only a fraction ($\sim 1/3$) of the annihilation energy goes into leptons \citep{hooper07, cumberbatch09}.  All dark matter models produce $\gamma = 2$ on average for $1~{\rm GeV} < E \ll M_{\rm DM}$.

\begin{figure}
\begin{center}
\epsfig{file=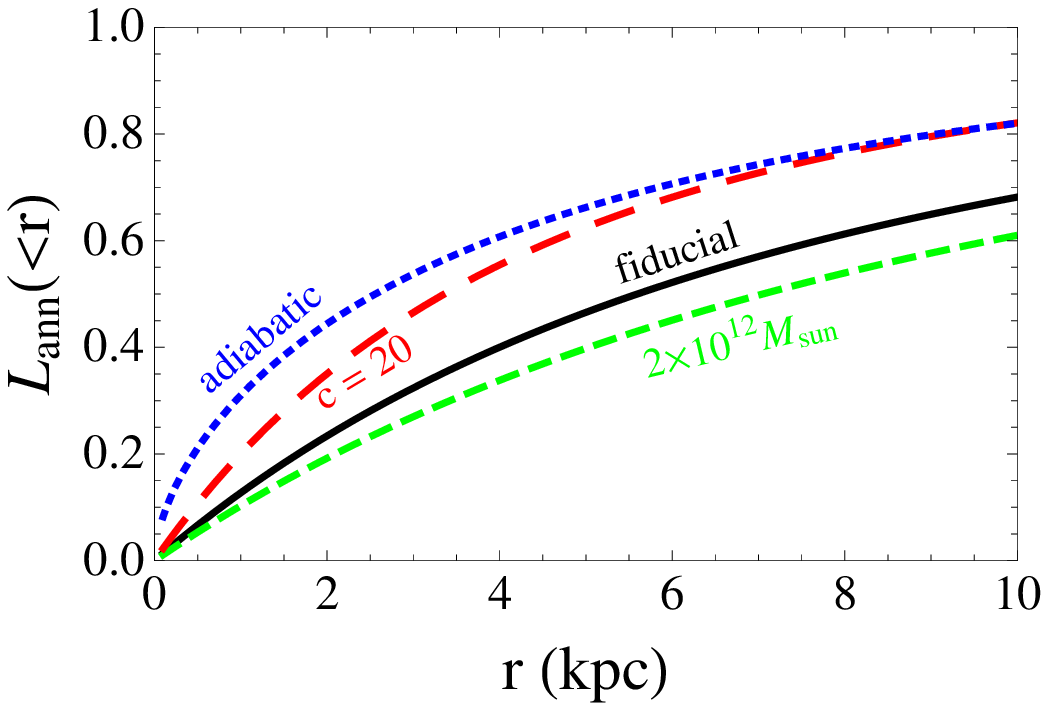, width=7cm}
\epsfig{file=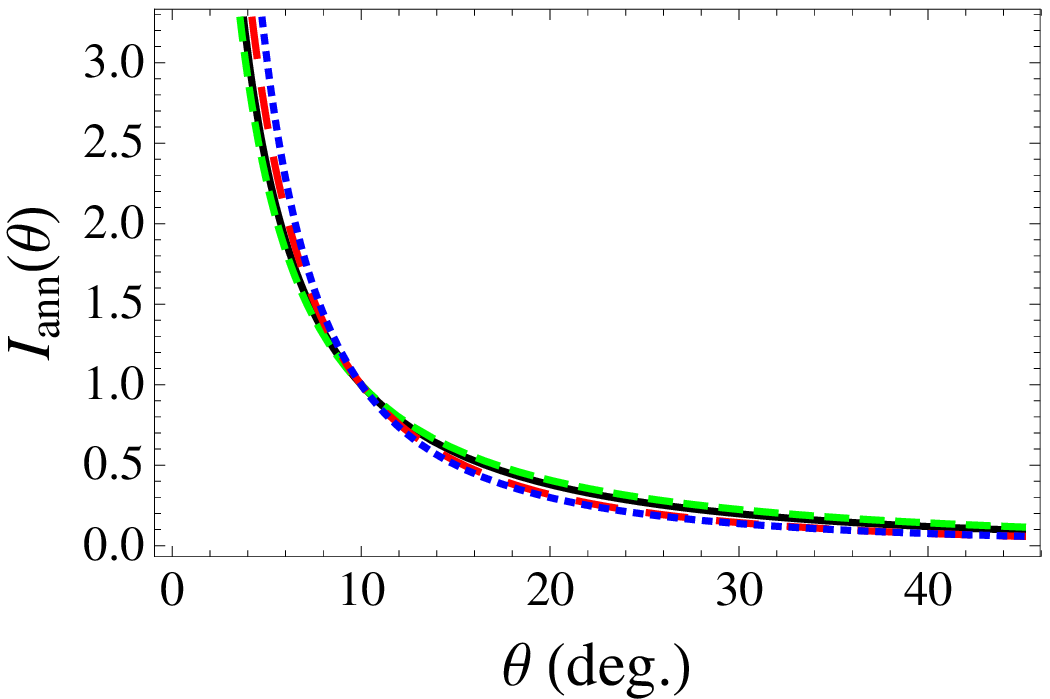, width=7cm}
\end{center}
\caption{Top Panel:
Cumulative dark matter annihilation luminosity in units of the total (assuming our fiducial parameters of $M_{\rm MW} = 10^{12}~M_{\odot}$, $c = 12$, and an NFW halo profile, unless stated otherwise).  The top curve labelled ``adiabatic'' instead assumes an inner power-law of $-1.2$.  The second curve down instead assumes $c = 20$.  The third down is our fiducial model, and the bottom curve instead assumes $M_{\rm DM} = 2\times 10^{12}~M_\odot$.    Bottom panel: The angular profile of the intensity in annihilations for the same four models.
\label{fig:ann_power}}
\end{figure}

\begin{figure}
\begin{center}
\epsfig{file=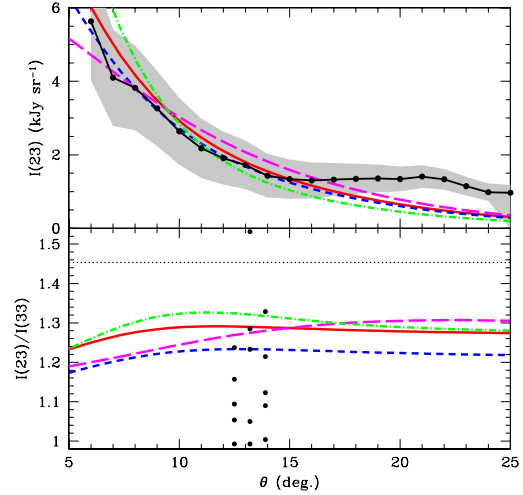, width=7cm}
\end{center}
\caption{Same as Figure \ref{fig:sn}, but instead for four dark matter models.  The red solid curve is a model with $M_{\rm DM} = 100$~GeV, $K_0 = 10^{28}$~cm$^{2}$~s$^{-1}$, $\delta =0.3$, and ${\cal B} = 10$.  The green dot-dashed curve is the same as the solid curve but with an inner power-law slope of $-1.2$, the blue short-dashed curve is the same as the solid curve but for a $10$ times more massive WIMP and ${\cal B} = 800$, and the magenta long-dashed curve is the same as the solid curve but with a $3$ times larger value for $K_0$.
\label{fig:wimpsync}}
\end{figure}

For our fiducial Milky Way halo model, the density profile scales as $r^{-1}$ out to $\approx 10$~kpc, before steepening.  As a result (assuming $d{\cal B}/dr = 0$), an equal amount of annihilation power per unit radius is emitted at all inner radii, and half of the annihilation power originates from within $5$~kpc for the fiducial halo model.  The top panel in Figure \ref{fig:ann_power} shows the cumulative energy output from annihilations as a function of radius for four plausible halo profiles.  The bottom panel in Figure \ref{fig:ann_power} shows what this profile would look like if the diffusion length were much less than the radial size of the annihilation profile and for a uniform $B_\perp$.  (This profile also corresponds to the intensity profile for direct annihilations into $\gamma$-rays.)  This figure suggests that the angular properties of the annihilation profile depend weakly on the dark matter halo model for plausible models (assuming that substructure is unimportant).

Figure \ref{fig:wimpsync} shows the haze radial intensity profile for four illustrative dark matter models in the $23~$GHz channel (left panel) and also the haze hardness (right panel), expressed as the ratio of the intensity in the $23~$GHz and $33~$GHz channels.  These models have been tuned to fit the observed radial profile, all with $\langle \sigma v \rangle = 3 \times 10^{-26}$~cm$^3$ s$^{-1}$.  The red solid curve assumes $M_{\rm DM} = 100$~GeV, $K_0 = 10^{28}$~cm$^{2}$~s$^{-1}$, $\delta =0.3$, and ${\cal B} = 10$.  The green dot-dashed curve is the same as the solid curve but with an inner power-law slope of $-1.2$ (what simulations typically find that include gas cooling; \citealt{tissera09}).  The blue short-dashed curve is the same as the solid curve but for a $10$ times more massive WIMP and ${\cal B} = 800$, and the magenta long-dashed curve is the same as the solid curve but for a value for $K_0$ that is $3$ times larger.  All models produce similar intensity profiles, and the hardness of the radiation does not depend strongly on angle because significant annihilations occur even at several kpc from the Galactic Center.  In detail, the less massive WIMP models produce a slightly softer spectrum, and all WIMP models are hardest within $r < \lambda$ from the Galactic Center.

It is simple to understand more quantitatively why the hardness of the electrons is not a strong function of radius if dark matter annihilations create the haze.  Ignoring the boundaries of the diffusive region, the number of electrons at a given point in our dark matter models is given by 
\begin{eqnarray}
n(E, r) &=& \frac{2 \, {\cal B}\, \langle\sigma v\rangle}{M_{\rm DM}^2} \int d^3r'  \, \rho(|\mathbf{r}+ \mathbf{r'}|)^2 \; G(r' | \lambda(E)), \\
&=& \frac{2 \, {\cal B} \,\langle\sigma v\rangle}{M_{\rm DM}^2} \int d^3r'  \frac{\rho_0^2 \,r_s^2}{|\mathbf{r}+ \mathbf{r'}|^2} \, G(r' | \lambda(E)), \\
  &=& \frac{2 \,{\cal B} \,\langle \sigma v \rangle \, \rho_0^2 \,r_s^2}{M_{\rm DM}^2 \, r^2}  \left( 1 + {\cal O}(\frac{\lambda^2}{r^2}) \right) ~~~~ \lambda \ll r,
\end{eqnarray}
where $\rho(r)$ is the dark matter halo profile, and the second line approximates the halo profile as the $r^{-1}$ inner cusp of an NFW profile with scale radius $r_s$.  The last line evaluates this expression in the limit  $\lambda \ll r$.  When $r > \lambda$, the hardness of the electron population or $n(E_1, r)/n(E_2, r)$ essentially does not depend on $r$.  This is in contrast to our other models, where the dependence of the hardness on $r$ is exponential when $r> \lambda$ because $n(E, r) \propto G(r | \lambda(E))$.  In the opposite regime where $\lambda > r$, all haze models have a hardness that is a weak function of radius.  The haze electrons cannot be in the latter regime at all radii and still reproduce the observed haze profile.

\section{Comparison of Models}
\label{sec:results}

\begin{table*}
\begin{center}
\caption{Parameters of the different haze models plotted in Figure \ref{fig:haze}.  All models assume $U_{\rm r} + U_{\rm B} = 10$~eV~cm$^{-3}$ and $B_\perp = \sqrt{2/3} \,10$~$\mu$G.  Models C--F assume that the input cosmic ray spectrum has sharp cutoffs at $1$ and $1000$~GeV.  The ``line-style'' column provides the color and line-style used in Figure \ref{fig:haze}. \label{table1}}
\begin{tabular}{l | c | c | c | c | c | c}
\hline  model & description & $\gamma_{\rm int}$ & $\dot{E}_{\rm tot}$ or $E_{\rm tot}$ (cgs) & $K_0$~(cm$^{-2}$~s$^{-1}$) & $\delta$ & line-style \\
\hline
A & $100$ GeV WIMP, ${\cal B} = 10$ & - & - & $10^{28}$ & $0.3$ & red solid\\
B & $1$ TeV WIMP, ${\cal B} = 800$ & - & - & $10^{28}$ & $0.3$ & green long dashed\\
C & GC Event, $1$ Myr ago & $2.5$ &$8\times10^{51}$ & $2 \times 10^{28}$ & $0.6$ & black dot-dashed\\
D & GC Event, $1$ Myr ago  & $2$ &$5\times10^{51}$ & $2 \times 10^{28}$ & $0.6$ & magenta dot-long dash\\
E & GC Population & $1.5$ &$ 10^{38}$ & $2\times 10^{28}$ & $0.6$ & brown dotted\\
F & GC Supernovae$^1$ & $2$ &$ 2\times 10^{38}$ & $2\times 10^{29}$ & $0.3$ & blue short-long dashed\\

\hline
\end{tabular}
\end{center}
$^1$ The soft synchrotron has been subtracted from this model as described in Section \ref{sec:population}.
\end{table*}

Table \ref{table1} summarizes the parameters of $6$ models for the haze emission that were motivated by the scenarios described in the previous section.  The parameters of these models have been selected to yield a decent fit to the \citet{dobler08} measurement at $23$ GHz in the inner $20^\circ$.  Models A and B are two dark matter models for the haze (Section \ref{sec:source_DM}), models C and D are models where it is due to a single event (Section \ref{sec:source_ps}), and Model F is where the haze is due to a Galactic Center population of supernovae (Section \ref{sec:SN}).  In Model E, an unknown Galactic Center population possessing a hard spectrum with $\gamma_{\rm int}= 1.5$ is responsible for the haze (Section \ref{ss:othercont}).

\begin{figure}
\epsfig{file=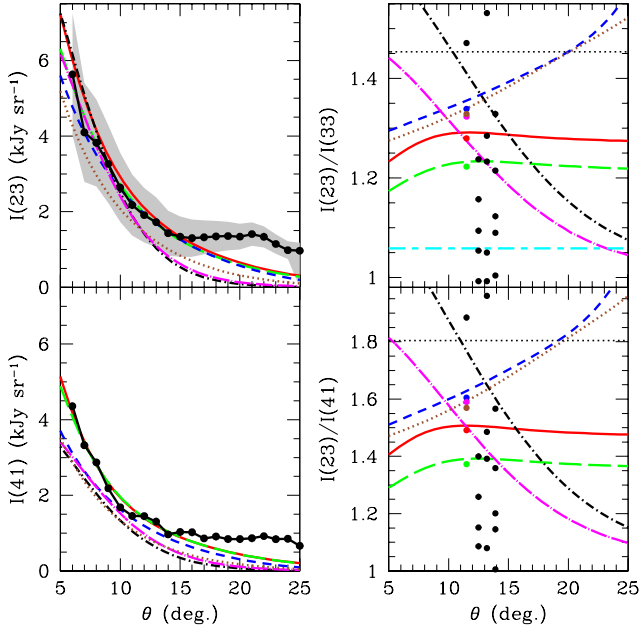, height=9cm}
\caption{Comparison of haze models described in Table 1 with the \citet{dobler08} measurement.  Left panels: The solid black curve with dots is the measured intensity profile of the WMAP haze in the $23$ and $41$ GHz channels.  The other curves are the intensity profile predicted by the different models.  Right panels:  The ratio of the intensity in the $23$ and $33$ GHz channels (top panel) and of the $23$ and $41$ GHz channels (bottom panel) for these models.  The curves in which the spectrum softens with radius represent the continuous source models, those in which this ratio is not a strong function of angle represent the dark matter models, and those that harden with radius represent the burst models.  The column of black dots represent the measured hardness of the different haze estimators as described in Figure 3.   For comparison, the leftmost column of colored dots at $11^\circ$ are the ratio of the model intensities in these two bands averaged over $6-15^\circ$, where colors are chosen to match the colors of the corresponding model curves.  Lastly, the cyan short-long dashed horizontal line is the intensity ratio expected for free-free emission, and the dotted black horizontal line is the intensity ratio of the galactic soft synchrotron radiation where $I_\nu \sim \nu^{-1}$.  \label{fig:haze}}
\end{figure}

Figure \ref{fig:haze} compares the microwave intensity profile (left panels) and microwave hardness (right panels) of the $6$ models against that of the haze measurement.  This comparison is the same as in Figure \ref{fig:sn} except for two differences.  First, the haze measurement in the $41$~GHz channel is also compared with the models (bottom panels).  The shaded region that brackets the estimate for the systematic errors is not included in the bottom left panel because these systematics are large at $41$~GHz.  Instead, only the profile from the fiducial CMB estimator in \citet{dobler08} is included.  Second, there are colored dots in the right panels that represent the mean hardness for the model curves with the corresponding color.  These colored dots are calculated by taking the ratio of the intensity averaged between $6^\circ$ and $15^\circ$ in the two channels. 


All of the considered model classes make different predictions for the haze hardness.
The qualitative trends for the hardness with radius are robust to the choice of diffusion and other model parameters.  These trends rely on just the fact that more energetic electrons both diffuse more quickly and radiate their energy faster than less energetic ones.  Models E and F (the continuous source models) predict that the spectrum should soften considerably with increasing angle, and Model F has the largest diffusion coefficient of the considered models.   Models C and D (the single burst models) predict the opposite,  that the spectrum should harden with radius.   Finally, models A and B (the dark matter models) predict that the spectral hardness should be almost constant as a function of angle.  


\begin{figure}
\begin{center}
\epsfig{file=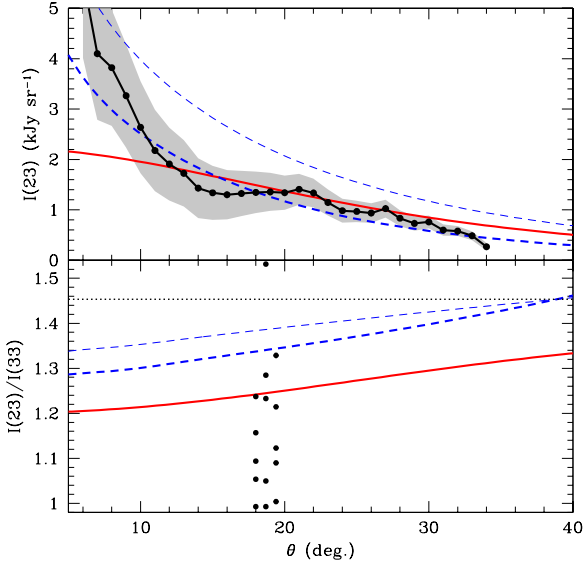, height=8cm}
\end{center}
\caption{Same as Figure \ref{fig:sn}, but for two models that have been tuned to fit the outskirts of the microwave haze emission.  These models use $A = 2 \times 10^{-16}~$GeV~s$^{-1}$ and $B_\perp = \sqrt{2/3}\;5~\mu$G, which are smaller than the fiducial values, and they both assume $\delta = 0.3$ and $L_{\rm haze} = 10~$kpc.  The thin blue dashed curves represent a supernova model with $\gamma_{\rm int} = 2$, $K_0 = 2 \times 10^{29}~$cm$^{-2}$~s$^{-1}$, and $\dot{E} = 3\times 10^{38}~$erg~s$^{-1}$.  The corresponding thick curves are the same, but subtracting off the soft-synchrotron component of this model as described in Section \ref{sec:population}.  The solid red curve represents a dark matter model with $K_0 = 6 \times 10^{28}~$cm$^{-2}$~s$^{-1}$, $M_{\rm DM} = 100~$GeV, and ${\cal B} = 8$.  \label{fig:outskirts}}
\end{figure}

\citet{dobler08} measured significant emission between $20^\circ$ and $35^\circ$ at the $1$~kJy~sr$^{-1}$ level.  All of our previous models for the haze emission begin to underpredict the emission compared to the haze measurement at $> 15^\circ$. There are claims that the haze at these large angles may not be significant \citep{cumberbatch09}.  If the fast decline in the haze intensity between $5^\circ$ and $10^\circ$ reflects the diffusion scale as we have so far assumed, it is difficult to explain the full extension of the haze with the previously considered models.  However, if the diffusion scale corresponds to the cutoff at $\sim35^\circ$ and the inner part of the haze owes to variation in the magnetic field, then models with larger values for the diffusion length may be able to explain this exterior haze emission.   Previous literature on the haze has taken different stances on whether to include in their fits the outer regions of the haze emission.  For example, \citet{hooper08} did not, whereas \citet{cholis08} did.   

Figure \ref{fig:outskirts} shows two simple models, a supernova and a dark matter model, in which the diffusion length has been increased to fit this exterior emission (by a factor of $2.2$ for the supernova model and $5.5$ for the dark matter model relative to the corresponding models in Table 1).  The models in this figure use $A = 2 \times 10^{-16}~$GeV~s$^{-1}$ and $B_\perp = \sqrt{2/3}\;5~\mu$G, which are smaller than our fiducial values, and they both assume $\delta = 0.3$ and $L_{\rm haze} = 10~$kpc.  These models do not attempt to fit the inner part of the haze, but allowing for a radial dependence of $B_\perp$ would likely provide the freedom for these models to accommodate the measured emission profile.  To fit the outer regions of the haze does \emph{not} require significantly more energy than in our previous models (despite the smaller $B$-field):  The supernova model assumes $\dot{E}_{\rm tot} = 3\times 10^{38}~$erg~s$^{-1}$ (a $50\%$ increase over Model F) and the dark matter model assumes ${\cal B} = 8$ (comparable to what is taken in Model A).  The angular dependence of the spectral hardness between these two cases is not as different as the cases in which $\lambda$ is smaller, but the predicted mean value is still much different between the dark matter and supernova models.

In all of our simplified models for the haze, we have made numerous simplifying assumptions.  For example, we fixed the value of $L_{\rm halo}$.  We find that changing $L_{\rm halo}$ does not significantly affect our conclusions because $L_{\rm halo} \gtrsim \lambda$ in all of our models.  We have also assumed a single magnetic field value.  The synchrotron specific intensity is proportional to the line-of-sight integral over $n_0 \, E^{3 - \gamma} \, B^2$.  Noting that at fixed frequency $E \propto B^{-1/2}$, this implies $I_\nu \sim B^{(1+\gamma)/2}$. 
If the Galactic $B$-field has a scale height of $2$~kpc, as is commonly assumed to fit radio data \citep{strong07}, this will act to suppress the emission at latitudes of $b \gtrsim 2 {\rm \, kpc}/8 {\rm \, kpc}~$rad.

\section{Inverse Compton Emission}
 \label{sec:IC} 
 
Inverse Compton scattering off of the hard electron population that is responsible for the haze emission can contribute to the galactic $\gamma$-ray emission (e.g., \citealt{hoopergray}).  An electron up-scatters via the inverse Compton process a photon of energy $E_{\rm seed}$ and wavelength $\lambda_{\rm seed}$ on average to an energy of
\begin{equation}
E_{\gamma} \approx 40 \, \left(\frac{\lambda_{\rm seed}}{1 \; {\rm \mu m}}\right)^{-1} \, 
\left(\frac{E}{80 ~{\rm GeV}} \right)^{2}~{\rm GeV},
\label{eqn:IC}
\end{equation}
where this equation is evaluated in the limit $\eta \equiv E_{\rm seed} E/(m_e c^2)^2 \ll 1$.  For $E$ and $\lambda_{\rm seed}$ of interest, this limit is not always strongly satisfied.  

Models for the interstellar radiation in the central kpc of the galaxy
find that starlight at $1^{+2}_{-0.7} \, \mu {\rm
m}$ contributes most of the energy density in radiation, while $\sim 10 \%$ is at $\lambda \approx 100~\mu$m from the repossessing of this light by dust \citep{moskalenko06}.  The majority of the emission that results in the $\gamma$-ray at $E_\gamma \approx 10 ~(50)$ GeV likely owes to the Compton scattering of starlight off of electrons with $E \approx 40~(90)$~GeV.  
These are slightly higher energy electrons than those that contribute to the microwave emission. Therefore, this $\gamma$-ray emission is sensitive to slightly higher energy electrons than the microwave synchrotron.  This statement assumes that $\gamma > 2$.  As $\gamma \rightarrow 2$, a comparable contribution to the $\gamma$-ray spectrum results from the inverse Compton scattering of $\sim 100~\mu$m dust emission on $10$ times more energetic electrons.

Interestingly, haze-like emission may have already been detected in the first year data from the Fermi Satellite in $10$s of GeV $\gamma$-rays \citep{doblerFermi}.  This emission was extracted in a similar manner to how the microwave haze was detected:  First, galactic templates for $\pi_0$ emission, for inverse Compton from the soft synchrotron electron population, and for a uniform extragalactic background were subtracted from the Fermi $\gamma$-ray sky map in different spectral bands.  Surprisingly, this resulted in a residual emission with similar spatial structure to the haze.  Specifically, \citet{doblerFermi} showed that this emission could be fit by a bivariate Gaussian with standard deviations of $15^\circ$ in longitude and $25^\circ$ in latitude and with central intensity $[E_\gamma I_E]_{\gamma} \approx 5 \times 10^{-7} \, {\rm GeV \,cm^{-2} \,s^{-1}\, sr^{-1}}$ at $40$~GeV.

We now investigate what the Fermi haze in combination with the microwave data reveals about the electron population in the haze region.  For a power-law electron distribution assuming $\eta \ll 1$, these measurements constrain the spectral index of these electrons to be
\begin{eqnarray}
\gamma &\approx& \bigg[2.7  - 0.9 \ln \left( \frac{[E_\gamma I_E]_\gamma}{10^{-6} \, {\rm GeV cm^{-2} s^{-1} sr^{-1}}}\, \frac{5 \, {\rm kJy\,sr}^{-1}}{[I_\nu]_{\rm 23 GHz}}\right)\nonumber \\
&-& 0.9 \ln \left( \frac{U_{\rm B}}{U_{\star}} \right)   + 1.3 \, \ln \left(\frac{E_{\gamma}}{30 \, \GeV} \,\frac{\lambda_{\rm seed}}{ 1\, {\rm \mu m}} \right) \bigg]\nonumber \\
&/& \left[1 + 0.4 \,\ln \left(\frac{E_{\gamma}}{30 \, \GeV} \,\frac{\lambda_{\rm seed}}{ 1\, {\rm \mu m}}\right) \right],
\label{eqn:gamma}
\end{eqnarray}
where we have assumed for simplicity $B_\perp = \sqrt{2/3} \; 10~\mu$G and $U_{\star}$ is the energy density in starlight, which we have assumed is concentrated at the wavelength of $\lambda_{\rm seed}$.\footnote{Equation~(\ref{eqn:gamma}) is derived using the relation
\begin{equation}
\frac{[E I_E]_R}{{[E I_E]_G}} = \frac{E_R^3 \, U_{\rm B} \, N(E_R)}{E_G^3 \, U_{\star} \, N(E_G)}, \nonumber
\end{equation}
where subscript $R$ is for the radio haze, $G$ is for the $\gamma$-ray, and $N(E) \propto E^{-\gamma}$ is the electron spectrum.  
}  However, we find that a more realistic spectral distribution has a minor effect on the inferred value of $\gamma$.  The logarithms evaluate to zero in equation~(\ref{eqn:gamma}) if values are taken for the Fermi and WMAP hazes that are consistent with measurements, yielding a soft spectrum with $\gamma \sim 2.7$.
Equation~(\ref{eqn:gamma}) primarily illustrates that there are significant astrophysical uncertainties inherent in the predictions of the $\gamma$-ray brightness of the the microwave haze electrons.  For example, a factor of $2$ uncertainty in $U_{\rm B}$ or $[E_\gamma I_E]_{\gamma}$ results in an uncertainty of $\approx 0.6$ in $\gamma$.

In addition, the inverse Compton spectrum alone can also be used to probe the spectrum of the electron population.  The measurement by \citet{doblerFermi} of the Fermi haze emission finds \emph{roughly}  $E_\gamma I_E \sim E^{-0.2}$ at $40$~GeV, with hints of a cutoff at higher energies.   Such a slope only requires $\gamma \approx 3.4$ (for $\eta \ll 1$) and is not evidence for a hard electron spectrum.

Different models for the microwave haze would produce different trends in the spectral hardness of their $\gamma$-ray emission.  If a single event produced the haze, it must have occurred $\Delta t < 4 \times 10^5 \, [10 \, {\rm eV \, cm}^{-3}/(U_{\rm r} + U_{\rm B})] \times [40 \,{\rm GeV}/E_\gamma]^{1/2}$~yr ago for energetic enough electrons to be present to emit at $E_\gamma$ for $\lambda_{\rm seed} = 1~\mu$m.  In addition, the $\gamma$-ray haze from an event would have a spectrum that hardens with radius, in contrast to the other haze models.  If instead the haze owes to a Galactic Center population (such as supernovae), the inverse Compton spectrum will be harder towards the Galactic Center again because the most energetic electrons would have diffused the shortest distances.  

For dark matter models, the hardness in the $\gamma$-ray emission should be fairly independent of radius from the Galactic Center because hard electrons are being produced in situ and do not have to diffuse from the Galactic Center.  Most WIMP dark matter models also produce tens of GeV $\gamma$-rays more directly as annihilation byproducts.  These direct byproducts should have an emission region that is less spatially extended than the inverse Compton emission from the haze electrons owing to diffusion effects.

\section{Conclusions}
We have investigated several models for the WMAP haze emission.  We argued that free-free emission is unlikely to be the source of this radiation, in agreement with previous studies.  Therefore, it is most likely that the haze is synchrotron.  If it is synchrotron, the haze indicates that the spectrum of electrons between a few and $20~$GeV is harder towards the Galactic Center.  

We showed that it is difficult to produce such a hardened source of Galactic Center electrons with the standard disk population of supernovae.  In addition, we argued that a disk population of pulsars (which has been cast by the literature as the astrophysical alternative to the dark matter annihilation model) is also unlikely to produce a hardened spectrum \emph{just} towards the Galactic Center.  The bulk of this paper investigated different models that may achieve sufficient hardening.  We showed that a Galactic Center population of supernovae and a diffusion-hardened electron distribution may be able to create the haze emission.  We found that $\sim 1000$ Galactic Center supernovae per Myr could be consistent with present constraints on the energetics, profile, and (marginally) the  spectral hardness of the haze if a relatively large value for the diffusion length is assumed.  This supernovae rate is consistent with predictions for its value in the central couple hundred parsecs.  

A transient event in the Galactic Center that releases $\sim 5\times 10^{51}$~erg in cosmic ray electrons in the past few Myr is another possibility that can reproduce the properties of the haze, even with a relatively soft spectrum ($\gamma_{\rm int} \approx 2- 2.5$).  It is unlikely that a single stellar event can source such energies, and, thus, such an event would be almost certainly associated with Sagittarius~A$^*$.  Intriguingly, there is evidence for enhanced star formation in the Galactic Center $\approx 2$~Myr ago.  This activity could have been accompanied by a brief episode of accretion onto the SMBH, potentially producing an accelerated population of electrons.  

Synchrotron radiation from the byproducts of dark matter annihilations is the most exciting explanation for the WMAP haze \citep{finkbeiner04, hooper08}. This explanation is naturally able to account for the measured microwave hardness of the WMAP haze emission and its radial extent.  This is true even if the diffusion length of $\sim 30$~GeV electrons is much smaller than the radius of the haze, in contrast to the other haze models.   A relatively light particle ($\lesssim 100$~GeV) is required in order not to invoke large boost factors in the annihilation rate beyond theoretically expected values.  However, even for a $100~$GeV WIMP, we require a boost factor of $\sim 10$ unless the inner halo density profile is steeper than a power-law with index $-1.2$.

Electrons with energies that radiate in the microwave and $\gamma$-ray are expected to lose significant energy as they diffuse a distance comparable to the extent of the haze. 
Because of how these losses occur in the different models, all of the considered models produce differing trends in the hardness of the microwave and $\gamma$-ray spectra as a function of angle from the Galactic Center.  These trends do not rely on the specific details of galactic cosmic ray propagation model.  Instead, all of these trends owe to the theoretically expected and empirically confirmed fact that more energetic particles diffuse more quickly and cool faster than less energetic ones.  Previous observations in the microwave and $\gamma$-ray are not yet able to constrain radial trends in the haze hardness, but future observations are projected to constrain them with high precision.  

We illustrated these trends using simple models that were tuned to be consistent with the \citet{dobler08} haze measurement.   In particular, dark matter annihilation models for the haze emission differ from astrophysical models in that a lot of the energy is input even at several kpc from the Galactic Center.  Because of this difference, the hardness of the synchrotron and inverse Compton emission in this model is not as strong a function of galactic latitude as in the astrophysical models.  This observable may allow one to distinguish this model from the others. 


Future observations have the potential to significantly constrain potential explanations for the WMAP haze.  In particular:
\begin{itemize}
\item Radio maps at frequencies higher than $408$~MHz would probe electrons that are more akin to the population that emits in the WMAP bands, and such maps could test the two electron population model for the microwave synchrotron emission.  In fact, such data already exists at $1420~$MHz \citep{2001A&A...376..861R}.
\item Planck temperature data (and possibly more sophisticated techniques applied to the WMAP data) can provide a better measurement of the haze profile as well as a measurement of the hardness as a function of angle. Planck's improved measurement relative to WMAP stems from the added channels at higher frequencies, which break degeneracies between the haze and other foregrounds \citep{dobler08}. 
\item Future CMB polarization measurements can search for a polarized haze.  While a recent analysis of WMAP polarization data did not detect a hard electron component \citep{gold10}, the haze should be present at some level if it owes to a hard population of synchrotron electrons.
\item $\gamma$-ray observations with Fermi are sensitive to slightly more energetic electrons than those that emit in the microwave.  Future $\gamma$-ray measurements have the potential to constrain the hardness of these electrons as a function of angle and thereby haze models.
\end{itemize}

We thank Greg Dobler for providing the data for the \citet{dobler08} measurement of the haze, and Lars Hernquist, Philip Hopkins, Eliot Quataert, Avi Loeb, Ryan O'Leary, and Jennifer Yeh for useful discussions.  We would especially like to thank Greg Dobler  and David Spergel for useful comments on the manuscript.  MM is supported by the NASA Einstein fellowship.
 \bibliography{dm_references}

\end{document}